%% file: Searchforglueball.tex
\documentclass{article}
\usepackage{arxiv}
\usepackage{xspace}
\usepackage[utf8]{inputenc} 
\usepackage[T1]{fontenc}    
\usepackage{hyperref}       
\usepackage{url}            
\usepackage{booktabs}       
\usepackage{amsfonts}       
\usepackage{nicefrac}       
\usepackage{microtype}      
\usepackage{graphicx}
\usepackage{lipsum}
\DeclareUnicodeCharacter{2212}{-}
\begin{document}
\input{commands.tex}
\title{Search for higher mass resonances via KK decay channel in pp collisions with ALICE at the LHC 
\thanks{Presented at Quark Matter 2022, 29$^{th}$ international conference on ultrarelativistic nucleus-nucleus collisions} }
\author{ Dukhishyam Mallick ( for the ALICE Collaboration) \\
	dukhishyam.mallick@cern.ch \\
	National Institute of Science Education and Research, HBNI, Jatni, Odisha, India
 }
{
\date{}
\maketitle
\begin{abstract}
	The quark model has proven successful in describing the basic building blocks of strongly 
	interacting particles in the Standard Model, where hadronic states consist of quarks and gluons.
	At the same time, Lattice QCD predicts the possibility of glueball candidates in the mass range
	\mbox{1550--1750} MeV/$c^2$, which have never been observed. The experimental search for the existence of
	mesons with no quark content is both interesting and challenging as the glueball is very likely to
	mix with surrounding quark-antiquark scalar meson states with the same quantum numbers. The
	large statistics data sample collected by ALICE in pp collisions at the highest LHC center-of-mass
	energy provides an opportunity to measure high mass resonances, whose characteristics and internal
	structure are still unknown. In this article, we report on the measurements of invariant mass distributions of higher mass resonances using the decay channels of K$^{0}_\mathrm{S}\mathrm{K}^{0}_\mathrm{S}$ and K$^{+}$K$^{-}$ pairs in pp collisions at $\sqrt{s}$ = 13 TeV using ALICE detector at midrapidity. 
\end{abstract}
\section{Introduction}
Quantum chromodynamics (QCD) is the theory of strong interaction between quarks mediated by gluons. QCD predicts that pairs or triplets of quark and antiquark can bind together, forming the hadrons. The strong force carriers are called gluons. In QCD, the gluons interact not only with the quarks but also among themselves, since they carry the color charge that characterizes the strong interaction. This fact allows Lattice QCD to predict for the existence of particles composed of gluons only~\cite{pdg}, the so called glueballs. The lightest glueball is expected in a mass range of 1550--1750 MeV/$c^{2}$ having total angular momentum (J), space parity (P), and charge parity (C), J$^{\mathrm{PC}}= 0^{++}$. The possible states with J$^{\mathrm{PC}}= 0^{++}$, and isospin I $=$0 are $f_\mathrm{0}(980)$, $f_\mathrm{0}(1370)$, $f_\mathrm{0}(1500)$ and $f_\mathrm{0}(1710)$. $f_\mathrm{0}(1710)$ is a suitable candidate for glueball as its mass is in the range expected by the Lattice QCD predictions~\cite{pdg, ks_prl}. ZEUS Collaboration observed states such as $f_{2}(1270)/a^{0}_{2}(1320)$,
f$^{\prime}_{2}(1525)$ and $f_\mathrm{0}(1710)$ in the mass spectrum of the decay channel K$^{0}_\mathrm{S}\mathrm{K}^{0}_\mathrm{S}$ pairs. 
It was found that the measured mass of $f_\mathrm{0}(1710)$ is consistent with the mass of glueball candidate and observed signal has a 5 standard deviation statistical significance~\cite{ks_prl}.
 \section{Description of detectors and analysis details}
The measurements have been performed using pp collisions at $\sqrt{s}$ = 13 TeV, Run 2 data collected by the ALICE detector. 
The total number of analysed events is about 1.5$\times$10$^{9}$. A detailed description of the ALICE detector setup and its performance is discussed in Ref.~\cite{alicedet}. The sub-detectors relevant for the studies are the Time Projection Chamber (TPC), the Time-of-Flight detector (TOF), the Inner Tracking System (ITS), covering pseudorapidity window of $|\eta| < $ 0.9,  and the  V0A (2.8 $ < \eta < $ 5.1) and  V0C (−3.7$ < \eta < $ −1.7) detectors. The resonances are reconstructed through the invariant mass distribution of inclusive K$^{0}_\mathrm{S}\mathrm{K}^{0}_\mathrm{S}$ and K$^{+}$K$^{-}$ pairs. \mbox{Fig.~\ref{fig:ks_sel}} shows the invariant mass distribution of K$^{0}_\mathrm{S}$ candidates reconstructed from $\pi^{+}\pi^{-}$ pairs. 
 \begin{figure}[htb]
\begin{center}
 \includegraphics[width =0.62\textwidth]{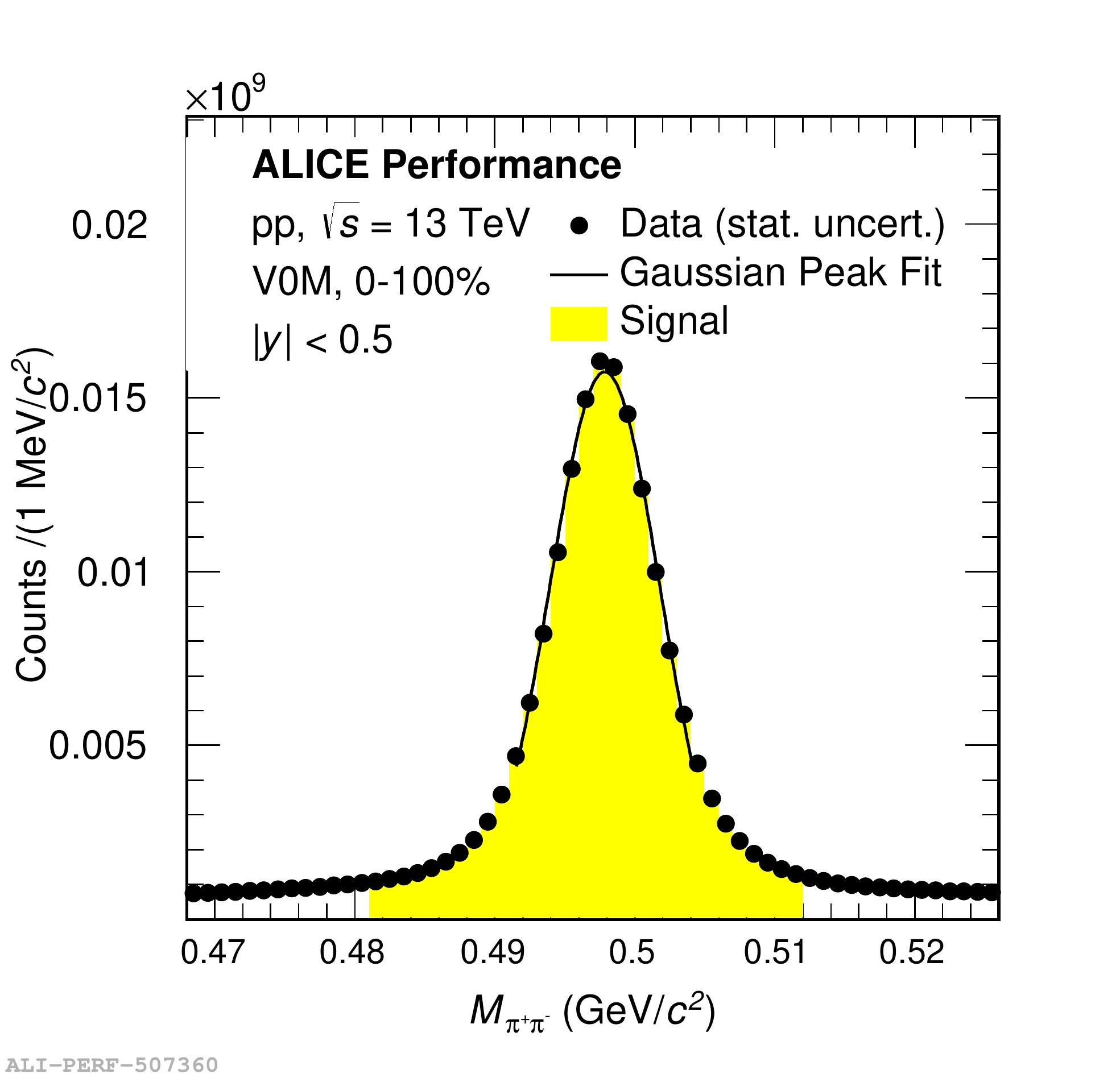}
 \caption{Invariant mass distribution of $\pi^{+}\pi^{-}$ pairs at midrapidity for 0--100$\%$  multiplicity class in pp collisions at $\sqrt{s}$ = 13 TeV. The yellow area represents the selected region of K$^{0}_\mathrm{S}$ invariant mass. To define the signal region, data have been fitted with a Gaussian function which is shown as black solid line.}     
	\label{fig:ks_sel}
  \end{center}
\end{figure}
For the reconstruction of K$^{0}_\mathrm{S}\mathrm{K}^{0}_\mathrm{S}$ invariant mass distributions, K$^{0}_\mathrm{S}$ candidates are selected in the mass window shown as yellow band in \mbox{Fig.~\ref{fig:ks_sel}} which covers 3 sigma area around the mass peak position. \mbox{Fig.~\ref{fig:inv_before}} shows the invariant mass distributions of K$^{0}_\mathrm{S}\mathrm{K}^{0}_\mathrm{S}$ (on the left) and  K$^{+}$K$^{-}$ (on the right) pairs before the combinatorial background subtraction (show as black markers) along with the combinatorial background distributions. The combinatorial backgrounds are reconstructed using mixed-event method for K$^{0}_\mathrm{S}\mathrm{K}^{0}_\mathrm{S}$ pairs, and like-sign method for K$^{+}$K$^{-}$ pairs. The mixed-event combinatorial background is normalized in the mass region 2.2 $<$ $M_{\mathrm{K}^{0}_\mathrm{S}\mathrm{K}^{0}_\mathrm{S}}$ $<$  2.3 GeV/$c^{2}$.
\begin{figure}[htb]
	\begin{center}
		\includegraphics[width =0.46\textwidth]{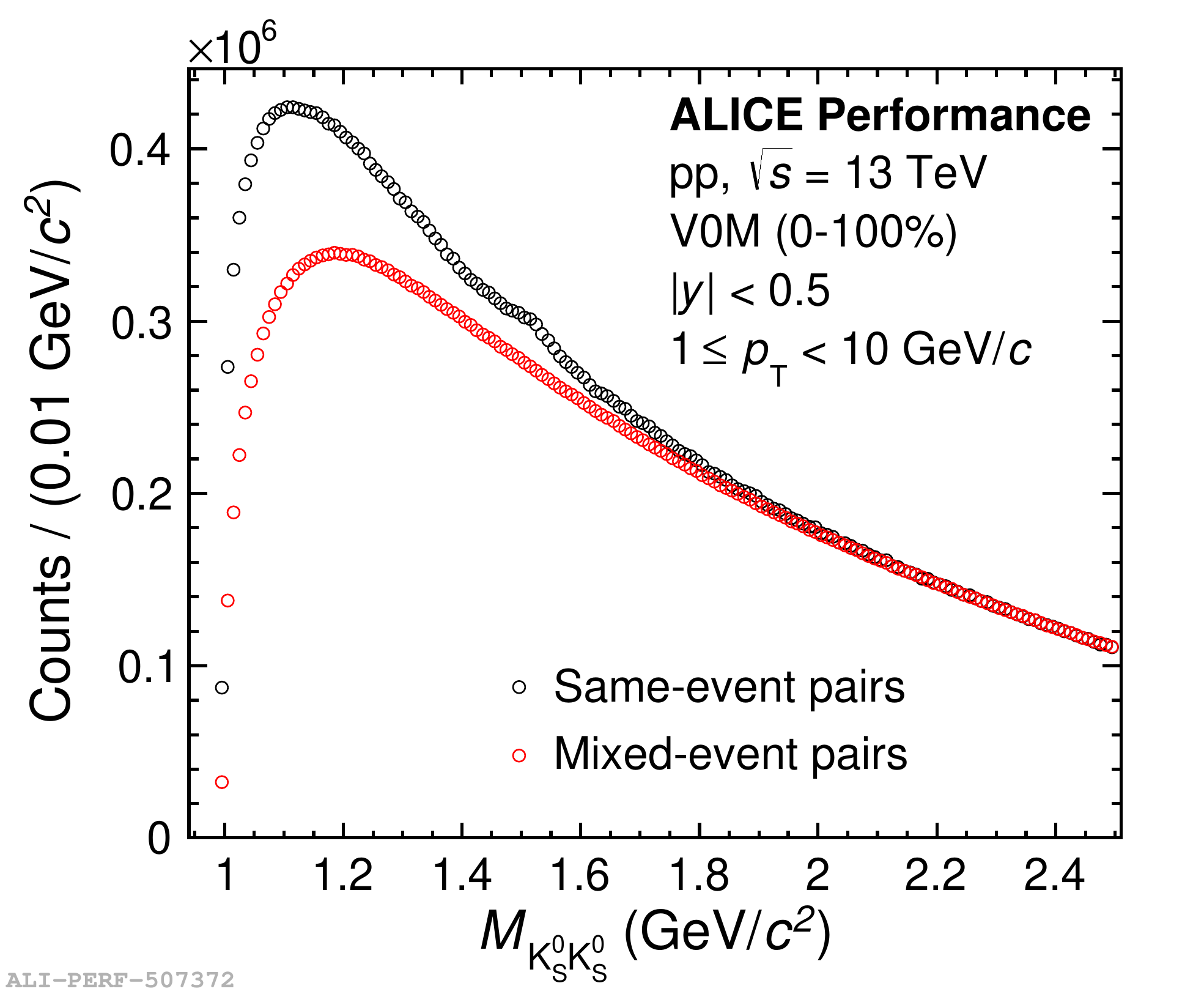}
		\includegraphics[width =0.46\textwidth]{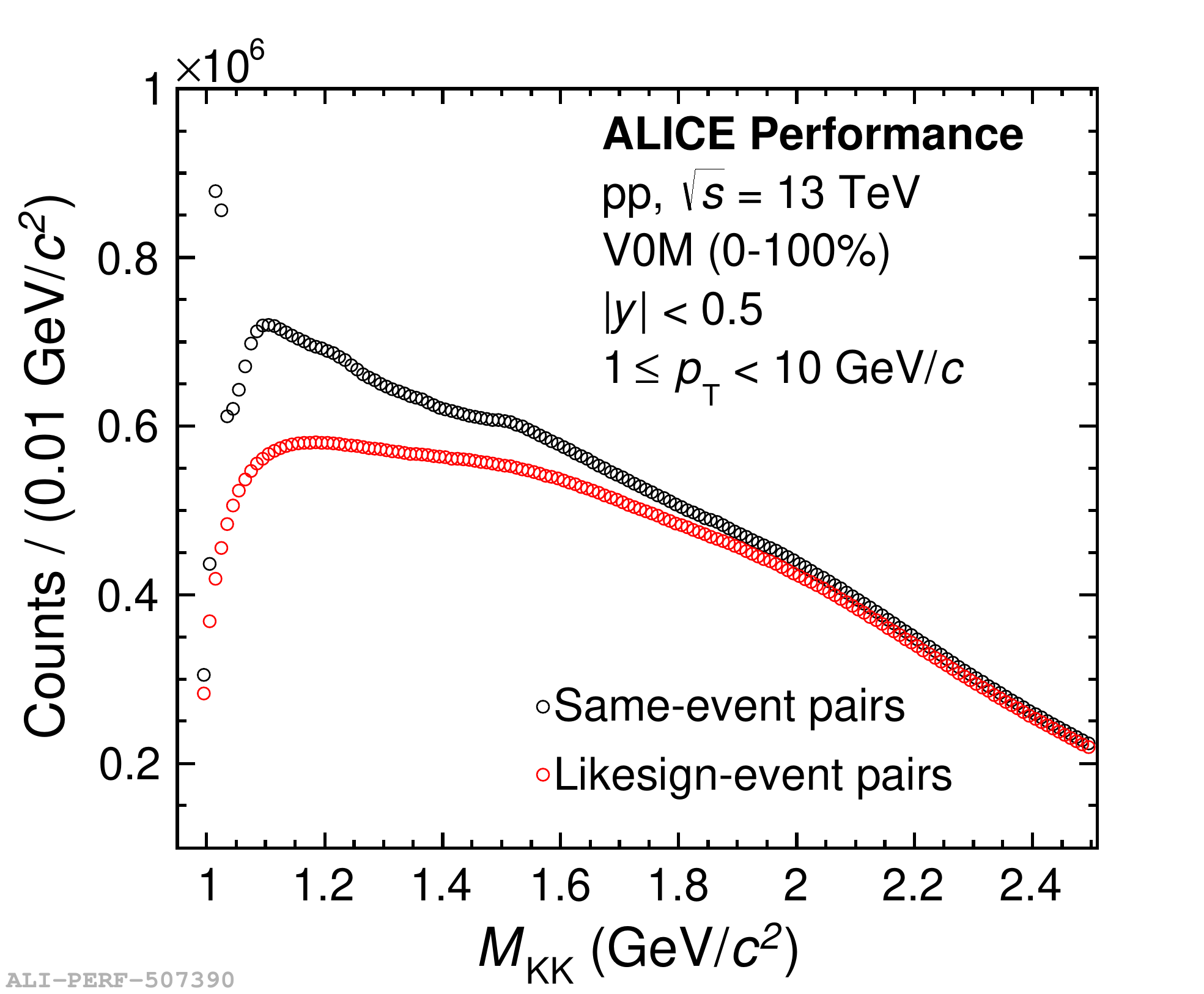}
	\end{center}
	\caption{The measured K$^{0}_\mathrm{S}\mathrm{K}^{0}_\mathrm{S}$ (left) and K$^{+}\mathrm{K}^{-}$ (right) invariant mass distributions before combinatorial background subtraction (black markers) and combinatorial background distributions (red markers).}     
	\label{fig:inv_before}
	\end{figure}
\section{Results and discussion}
 Figure~\ref{fig:inv_ks_after} shows the K$^{0}_\mathrm{S}\mathrm{K}^{0}_\mathrm{S}$ and K$^{+}$K$^{-}$ invariant mass distributions after combinatorial background subtraction in the multiplicity class 0--100$\%$ and transverse momentum range 1.0 $\leq \pt <$ 10 GeV$/c$ for pp collisions at $\sqrt{s}$ = 13 TeV. Multiple resonances appear in the mass spectrum and each of the resonance (R) distribution can be described by a relativistic Breit-Wigner (rBW) function
\begin{equation}
rBW(R) = \frac{M_{R}\Gamma_{0}M_{0} } {(M^{2}_{R} -M^{2}_{0})^{2} + M^{2}_{0}\Gamma_{0}^{2}}.   
\label{rBW}
\end{equation}
The invariant mass distribution of K$^{0}_\mathrm{S}\mathrm{K}^{0}_\mathrm{S}$ pairs is fitted with a sum of relativistic Breit-Wigner function (F(M)) used to describe signal of the resonances plus a smoothly varying background function (U(R)), motivated by SU(3) symmetry~\cite{su3}   
\begin{eqnarray}\nonumber 
F(M) &=&  a|5rBW(f_{2}(1270))- 3rBW(a^{0}_{2}(1320)) + 2rBW(f_{2}^{\prime}(1525))|^{2} 
\\    && +b |rBW(f_{0}(1710))|^{2}     + \nonumber
\\ 
U(R) &=&  A (M_{R} - 2m_{0})^{B} Exp(-C(M_{R} - 2m_{0})) ,
\label{Coherent_rBW}
\end{eqnarray}
where $M_\mathrm{R}$ is the mass of decay pairs of resonances, $M_\mathrm{0}$ and $\Gamma_{0}$ are mass and width of resonances, $m_\mathrm{0}$ is PDG mass of the decay daughter of the resonance, $\textit{a, b, c, A, B}$ and $\textit{C}$ are free parameters. 
A non-coherent sum of 3 relativistic Breit-Wigner functions plus U(R) function is used for describing the invariant mass distribution of K$^{+}$K$^{-}$ pairs. Three peaks are seen at mass around 1300, 1500, and 1700 MeV/$c^{2}$ in K$^{0}_\mathrm{S}\mathrm{K}^{0}_\mathrm{S}$ invariant mass distribution as shown the left panel of Fig.~\ref{fig:inv_ks_after}.
\begin{figure}[htb] 
	\begin{center}
		\includegraphics[width =0.48\textwidth]{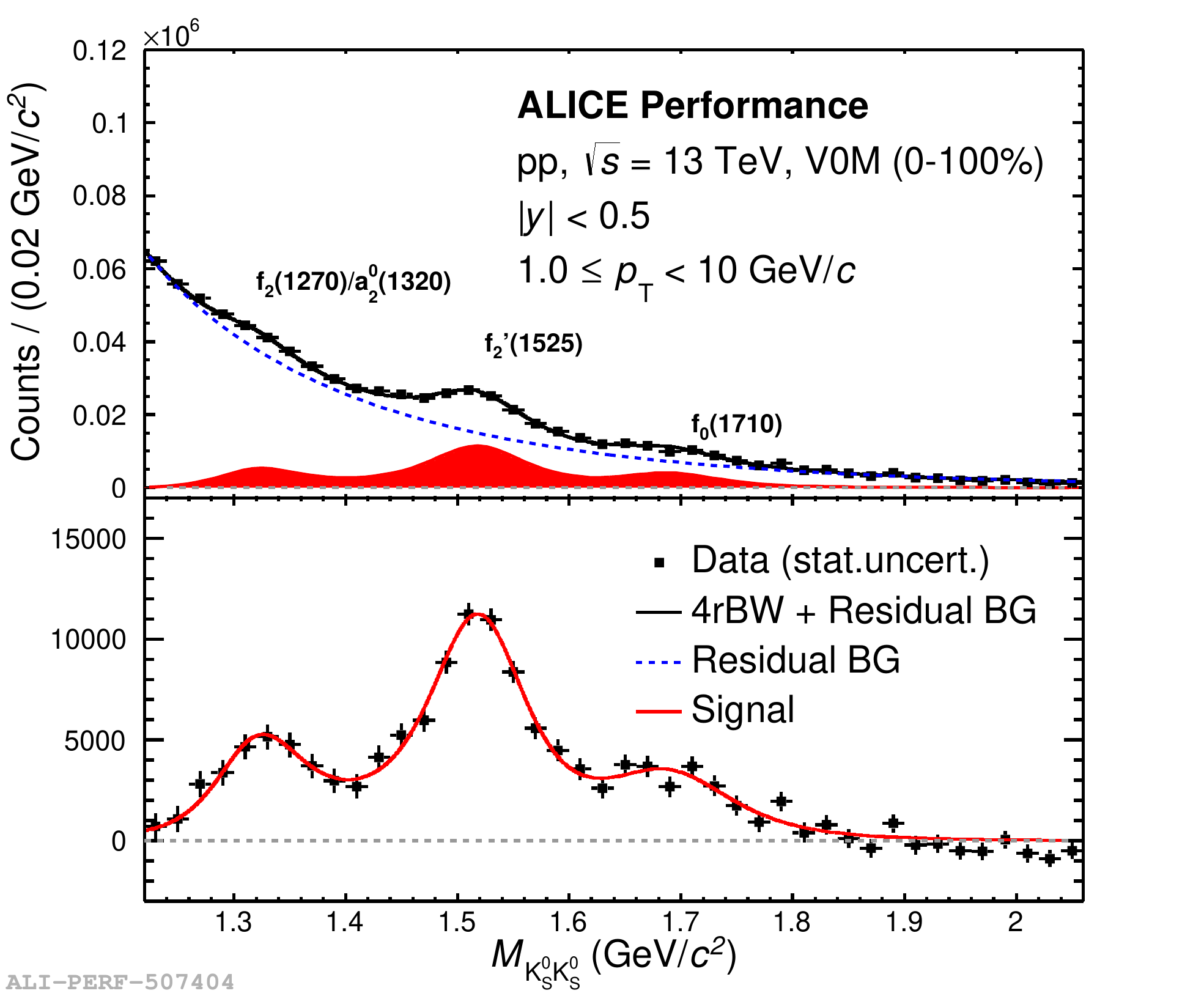}
		\includegraphics[width=0.46\textwidth]{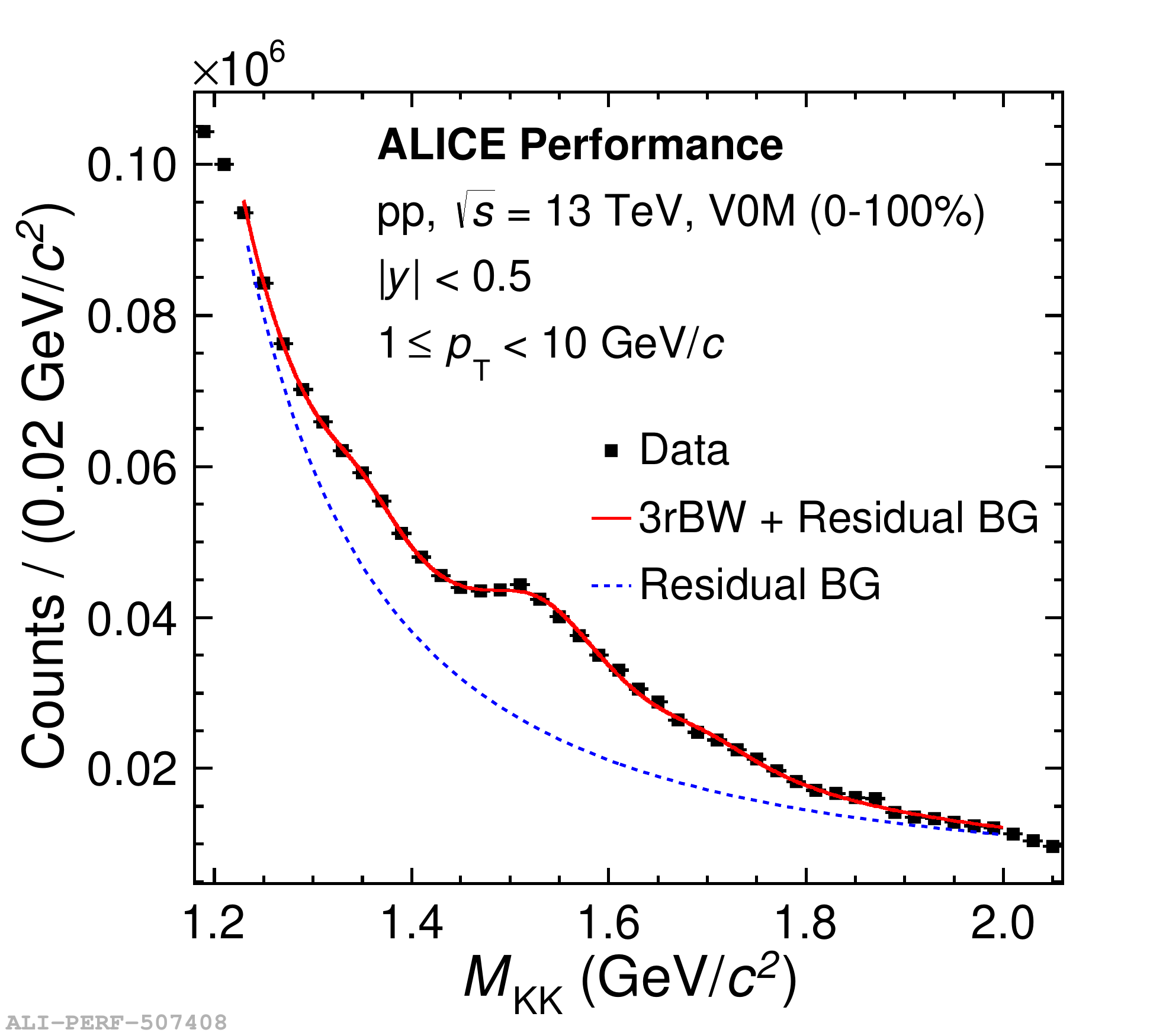}
	\end{center}
		\caption{Left: invariant mass distribution of K$^{0}_\mathrm{S}\mathrm{K}^{0}_\mathrm{S}$ pairs, 
		the black solid line is result of fit and the blue dashed line represents the residual background function described in the Eq.~\ref{Coherent_rBW}. Bottom panel shows the invariant mass distribution of K$^{0}_\mathrm{S}\mathrm{K}^{0}_\mathrm{S}$ pairs after  subtraction of the residual background, fitted with a sum of relativistic Breit-Wigner functions as described in the text. Right: measured K$^{+}\mathrm{K}^{-}$ invariant mass distribution. The solid line is fit function for resonances and dashed line represents residual background function.} 
	\label{fig:inv_ks_after}
\end{figure}
A prominent signal of resonance state $f_{2}^{\prime}(1525)$ is seen in mass spectrum of K$^{0}_\mathrm{S}$K$^{0}_\mathrm{S}$ and  K$^{+}$K$^{-}$ pairs. 
 \section{Summary}
We have reported the invariant mass distributions of higher mass resonances using decay of K$^{0}_\mathrm{S}$K$^{0}_\mathrm{S}$ and K$^{+}$K$^{-}$ in pp collisions at $\sqrt{s}$ = 13 TeV rregistered by the ALICE detector. In K$^{0}_\mathrm{S}\mathrm{K}^{0}_\mathrm{S}$ mass spectrum, 3 resonance states are observed in the mass range 1--2 GeV/$c^{2}$. The mass peak around 1710 MeV/$c^{2}$ corresponds to the mass of the expected glueball candidate with 0$^{++}$. A prominent mass peak around 1525 MeV/$c^{2}$ is observed in both K$^{0}_\mathrm{S}$K$^{0}_\mathrm{S}$ and  K$^{+}$K$^{-}$ mass spectrum. The high statistics data of upcoming Run 3 and Run 4 will provide further precision on these measurements.
\bibliography{searchforglueball}
\end{document}

%% file: commands.tex
%

\newcommand{\pp}           {pp\xspace}
\newcommand{\ppbar}        {\mbox{$\mathrm {p\overline{p}}$}\xspace}
\newcommand{\XeXe}         {\mbox{Xe--Xe}\xspace}
\newcommand{\PbPb}         {\mbox{Pb--Pb}\xspace}
\newcommand{\pA}           {\mbox{pA}\xspace}
\newcommand{\pPb}          {\mbox{p--Pb}\xspace}
\newcommand{\AuAu}         {\mbox{Au--Au}\xspace}
\newcommand{\dAu}          {\mbox{d--Au}\xspace}
\newcommand{\CuCu}         {\mbox{Cu--Cu}\xspace}

\newcommand{\s}            {\ensuremath{\sqrt{s}}\xspace}
\newcommand{\snn}          {\ensuremath{\sqrt{s_{\mathrm{NN}}}}\xspace}
\newcommand{\pt}           {\ensuremath{p_{\rm T}}\xspace}
\newcommand{\meanpt}       {$\langle p_{\mathrm{T}}\rangle$\xspace}
\newcommand{\ycms}         {\ensuremath{y_{\rm CMS}}\xspace}
\newcommand{\ylab}         {\ensuremath{y_{\rm lab}}\xspace}
\newcommand{\etarange}[1]  {\mbox{$\left | \eta \right |~<~#1$}}
\newcommand{\yrange}[1]    {\mbox{$\left | y \right |~<$~0.5}}
\newcommand{\dndy}         {\ensuremath{\mathrm{d}N_\mathrm{ch}/\mathrm{d}y}\xspace}
\newcommand{\dndeta}       {\ensuremath{\mathrm{d}N_\mathrm{ch}/\mathrm{d}\eta}\xspace}
\newcommand{\avdndeta}     {\ensuremath{\langle\dndeta\rangle}\xspace}
\newcommand{\dNdy}           {\ensuremath{\mathrm{d}N_\mathrm{ch}/\mathrm{d}y}\xspace}
\newcommand{\dNdyy}         {\ensuremath{\mathrm{d}N/\mathrm{d}y}\xspace}
\newcommand{\Npart}        {\ensuremath{N_\mathrm{part}}\xspace}
\newcommand{\Ncoll}        {\ensuremath{N_\mathrm{coll}}\xspace}
\newcommand{\dEdx}         {\ensuremath{\textrm{d}E/\textrm{d}x}\xspace}
\newcommand{\RpPb}         {\ensuremath{R_{\rm pPb}}\xspace}
\newcommand{\RAA}         {\ensuremath{R_{\rm AA}}\xspace}

\newcommand{\nineH}        {$\sqrt{s}~=~0.9$~Te\kern-.1emV\xspace}
\newcommand{\seven}        {$\sqrt{s}~=~7$~Te\kern-.1emV\xspace}
\newcommand{\twoH}         {$\sqrt{s}~=~0.2$~Te\kern-.1emV\xspace}
\newcommand{\twosevensix}  {$\sqrt{s}~=~2.76$~Te\kern-.1emV\xspace}
\newcommand{\five}         {$\sqrt{s}~=~5.02$~Te\kern-.1emV\xspace}
\newcommand{\twosevensixnn}{$\sqrt{s_{\mathrm{NN}}}~=~2.76$~Te\kern-.1emV\xspace}
\newcommand{\fivenn}       {$\sqrt{s_{\mathrm{NN}}}~=~5.02$~Te\kern-.1emV\xspace}
\newcommand{\LT}           {L{\'e}vy-Tsallis\xspace}
\newcommand{\GeVc}         {Ge\kern-.1emV/$c$\xspace}
\newcommand{\MeVc}         {Me\kern-.1emV/$c$\xspace}
\newcommand{\TeV}          {Te\kern-.1emV\xspace}
\newcommand{\GeV}          {Ge\kern-.1emV\xspace}
\newcommand{\MeV}          {Me\kern-.1emV\xspace}
\newcommand{\GeVmass}      {Ge\kern-.2emV/$c^2$\xspace}
\newcommand{\MeVmass}      {Me\kern-.2emV/$c^2$\xspace}
\newcommand{\lumi}         {\ensuremath{\mathcal{L}}\xspace}

\newcommand{\ITS}          {\rm{ITS}\xspace}
\newcommand{\TOF}          {\rm{TOF}\xspace}
\newcommand{\ZDC}          {\rm{ZDC}\xspace}
\newcommand{\ZDCs}         {\rm{ZDCs}\xspace}
\newcommand{\ZNA}          {\rm{ZNA}\xspace}
\newcommand{\ZNC}          {\rm{ZNC}\xspace}
\newcommand{\SPD}          {\rm{SPD}\xspace}
\newcommand{\SDD}          {\rm{SDD}\xspace}
\newcommand{\SSD}          {\rm{SSD}\xspace}
\newcommand{\TPC}          {\rm{TPC}\xspace}
\newcommand{\TRD}          {\rm{TRD}\xspace}
\newcommand{\VZERO}        {\rm{V0}\xspace}
\newcommand{\VZEROA}       {\rm{V0A}\xspace}
\newcommand{\VZEROC}       {\rm{V0C}\xspace}
\newcommand{\Vdecay} 	   {\ensuremath{V^{0}}\xspace}

\newcommand{\ee}           {\ensuremath{e^{+}e^{-}}} 
\newcommand{\pip}          {\ensuremath{\pi^{+}}\xspace}
\newcommand{\pim}          {\ensuremath{\pi^{-}}\xspace}
\newcommand{\kap}          {\ensuremath{\rm{K}^{+}}\xspace}
\newcommand{\kam}          {\ensuremath{\rm{K}^{-}}\xspace}
\newcommand{\pbar}         {\ensuremath{\rm\overline{p}}\xspace}
\newcommand{\kzero}        {\ensuremath{{\rm K}^{0}_{\rm{S}}}\xspace}
\newcommand{\lmb}          {\ensuremath{\Lambda}\xspace}
\newcommand{\almb}         {\ensuremath{\overline{\Lambda}}\xspace}
\newcommand{\Om}           {\ensuremath{\Omega^-}\xspace}
\newcommand{\Mo}           {\ensuremath{\overline{\Omega}^+}\xspace}
\newcommand{\X}            {\ensuremath{\Xi^-}\xspace}
\newcommand{\Ix}           {\ensuremath{\overline{\Xi}^+}\xspace}
\newcommand{\Xis}          {\ensuremath{\Xi^{\pm}}\xspace}
\newcommand{\Oms}          {\ensuremath{\Omega^{\pm}}\xspace}
\newcommand{\degree}       {\ensuremath{^{\rm o}}\xspace}
\newcommand{\kstar}        {\ensuremath{\rm {K}^{\rm{* 0}}}\xspace}
\newcommand{\phim}        {\ensuremath{\phi}\xspace}
\newcommand{\pik}          {\ensuremath{\pi\rm{K}}\xspace}
\newcommand{\kk}          {\ensuremath{\rm{K}\rm{K}}\xspace}
\newcommand{\kskm}{$\mathrm{K^{*0}/K^{-}}$}
\newcommand{\phikm}{$\mathrm{\phi/K^{-}}$}
\newcommand{\phixi}{$\mathrm{\phi/\Xi}$}
\newcommand{\phiom}{$\mathrm{\phi/\Omega}$}
\newcommand{\xiphi}{$\mathrm{\Xi/\phi}$}
\newcommand{\omphi}{$\mathrm{\Omega/\phi}$}
\newcommand{\kstf} {K$^{*}(892)^{0}~$}
\newcommand{\phf} {$\mathrm{\phi(1020)}~$}
\newcommand{\dd}{\ensuremath{\mathrm{d}}}
\newcommand{\mT}{\ensuremath{m_{\mathrm{T}}}\xspace}
\newcommand{\krr}{\ensuremath{\kern-0.09em}}

%% file: Searchforglueball.bbl
\begin{thebibliography}{1}
\bibitem{pdg} R. L. Workman et al., PTEP 2022 (2022) 083C01. 	
\bibitem{ks_prl} ZEUS Collaboration, S. Chekanov et al., PRL 101,112003 (2008).
\bibitem{alicedet} K. Aamodt et al. [ALICE], JINST 3, S08002 (2008). 	
\bibitem{su3} D. Faiman, H.J. Lipkin, H.R. Rubinstein, Phys. Lett. B 59 (1975) 269-273.
\end{thebibliography}
